\documentclass[12pt,psfig,a4,twoside]{article}
\usepackage{geometry}
\usepackage{graphicx}
\usepackage{amsmath}
\usepackage{graphics}
\usepackage{color}
\usepackage{setspace}
\usepackage{mathrsfs}
\usepackage{amsbsy,amssymb,amsmath,ulem}
\usepackage{dcolumn}
\usepackage{bm}
\usepackage{hyperref}
\hypersetup{colorlinks=true,
			linkcolor=magenta,
			citecolor=blue}
\usepackage{fancyhdr}
\fancypagestyle{firstpagestyle}
{
	\fancyhf{}
	\fancyhead[L]{\it\footnotesize Under consideration in Continuum Mech. and Thermodynamics}
	\fancyhead[R]{\footnotesize \thepage}
}

\newcommand{\PSI}{{\boldsymbol{\Psi}}}
\newcommand{\scU}{{\mathscr{U}}}

\newcommand{\scE}{{\mathscr{E}}}
\newcommand{\calI}{\mathcal{I}}
\newcommand{\calA}{\mathcal{A}}

\newcommand{\BV}{{\boldsymbol{V}}}

\newcommand{\DD}{{\boldsymbol{D}}}

\newcommand{\EE}{{\boldsymbol{E}}}
\newcommand{\FF}{{\boldsymbol{F}}}
\newcommand{\AAA}{{\boldsymbol{A}}}

\newcommand{\UU}{{\boldsymbol{U}}}

\newcommand{\ww}{{\boldsymbol{w}}}
\newcommand{\xx}{{\boldsymbol{x}}}

\newcommand{\BB}{{\boldsymbol{B}}}
\newcommand{\GG}{{\boldsymbol{G}}}

\newcommand{\qq}{{\boldsymbol{q}}}

\newcommand{\II}{{\boldsymbol{I}}}
\newcommand{\TT}{{\boldsymbol{T}}}

\newcommand{\pp}{{\boldsymbol{p}}}
\newcommand{\vv}{{\boldsymbol{v}}}

\newcounter{mypar}
\addtocounter{mypar}{0}

\begin{document}

\small

\title{\bf\large A hyperbolic model for viscous Newtonian flows
}


\author{
	Ilya Peshkov\thanks{Aix Marseille Universit\'{e}, CNRS, IUSTI UMR 7343, Marseille, France. On leave from Sobolev Institute of Mathematics, Russian Academy of Sciences, Novosibirsk, Russia; \href{mailto:peshkov@math.nsc.ru}{\nolinkurl{peshkov@math.nsc.ru}}}, \ \ \ 
	Evgeniy Romenski\thanks{Sobolev Institute of Mathematics, Russian Academy of Sciences, Novosibirsk, Russia; \href{mailto:evrom@math.nsc.ru}{\nolinkurl{evrom@math.nsc.ru}}}
	\vspace{20pt}
}

\date{}
\maketitle

\maketitle
\thispagestyle{firstpagestyle}
\begin{abstract}
We discuss a pure hyperbolic alternative to the Navier-Stokes equations, which are of parabolic type. As a result of the substitution of the concept of the viscosity coefficient by a microphysics-based temporal characteristic, particle settled life (PSL) time, it becomes possible to formulate a model for viscous fluids in a form of first order hyperbolic partial differential equations. Moreover, the concept of PSL time allows the use of the same model for flows of viscous fluids (Newtonian or non-Newtonian) as well as irreversible deformation of solids. In the theory presented, a continuum is interpreted as a system of  material particles connected by bonds; the internal resistance to flow is interpreted as elastic stretching of the particle bonds; and a flow is a result of bond destructions and rearrangements of particles. Finally, we examine the model for simple shear flows, arbitrary incompressible and compressible flows of Newtonian fluids and demonstrate that Newton's viscous law can be obtained in the framework of the developed hyperbolic theory as a steady-state limit. A basic relation between the viscosity coefficient, PSL time, and the shear sound velocity is also obtained. 

\end{abstract}

\hypersetup{linkcolor=black}
\tableofcontents

\hypersetup{linkcolor=red}

\section{\label{Intro}Introduction}

\subsection{\label{sec:solidsfluids}Solids vs. fluids}

Under normal conditions, solids and fluids behave differently; however, in this paper, we demonstrate that from the continuum physics point of view such differences are of a quantitative character and that the dynamics of both these states of matter can be modeled with a single system of partial differential equations (PDEs). In particular, we demonstrate that there is a microphysics-based alternative to the conventional concept of viscosity in its traditional and phenomenological meaning, \textit{i.e.}, as a coefficient of proportionality between the strain rate and stress tensor. With such an alternative, it becomes possible to describe flows of viscous fluids, whether Newtonian or non-Newtonian, from the same basis as irreversible deformation in solids. 

Let us first describe an approximate physical model underlying the continuum theory discussed below. 
We start the discussion from the central assumption of any continuum theory~\cite{LandauLifshitzHydro}, that is solids and fluids are represented by a disordered system of material particles, \textit{i.e.}, infinitesimal control volumes, with a constant mass. These particles must be small enough to be considered as infinitesimal volumes if compared with the entire system. At the same time, the material particles must contain enough real molecules or atoms in order to volume averaged state parameters as the mass density $\rho$, momentum $\rho\vv$, and entropy density $\rho s$  are well-defined. 

In addition to these conventional state parameters, we assume that the particles possess one more degree of freedom. Namely, we assume that the particles, in fact small volumes, are deformable and can rotate. The deformation and rotation of a particle can be characterized by a non-symmetric $3\times3$-matrix $\AAA$, called here distortion (or more precisely elastic distortion~\cite{God1978,GodRom2003}). We first of all think of $\AAA$ as a local field, \textit{i.e.}, the distribution of $\AAA$ does not related to the global deformation\footnote{The distribution of $\AAA$ gives the global deformation of the continuum if flow is absent, \textit{i.e.}, the deformation is reversible, which is not the case in this study.} of the medium, but describes the shape and orientation of the particles solely.

Moreover, in order to have changes in $\AAA$, we should assume that the material particles are able to interact with each other. More precisely, we assume that the particles are connected by bonds. The notion of the bond between two neighboring material particles (small volumes) has a meaning that the particles are in contact with each other, \textit{i.e.}, they have a common boundary (interface), which they have a tendency to retain. Of course, the reasons for that lie in a more microscopic scale than the scale of material particles. It is the diffusion mass flux through the common interface, which due to inertia plays the role of a bond resistant to shearing of particles in the case of gases, or it can be  real chemical bonds between molecules located along the both sides of the interface in the case of solids or fluids in a condensed state (liquids).
Thus, we stipulate for the further discussion that when we talk about elastic stretching of the bond between two particles, we mean that these particles deform while staying in contact with each other via the common interface. In turn, the bond destruction means that the two particles have lost the direct contact.

With this physical model in mind, it is natural to question how such a system of particles can flow, \textit{i.e.}, deform irreversibly. Clearly, the system can flow only if the bonds between particles can be destroyed; if the bonds exist forever, the material behaves elastically, which is not the case in this study. If the bonds can be  destroyed, then a second question arises: what is a characteristic time $\tau$ of the existence of particle bonds while the system flows, \textit{i.e.}, the time after which the bonds are destroyed? Clearly, $\tau$ cannot be zero because that would mean that no bonds exist. Thus, the time $\tau$ is finite, with $0<\tau<\infty$, which rises the third basic question of what the particle bonds do while the system deforms during time $t<\tau$? As we assume that bonds are stretchable, \textit{i.e.}, material particles are deformable, otherwise the system cannot deform, the answer is that bonds are stretched elastically because bond destruction is forbidden at time~$t<\tau$.

Therefore, it becomes clear that the true source of irreversible deformation in a system of connected material particles is the process of particle rearrangement. Hence, this  can be decomposed into three basic subprocesses: elastic bond stretching, bond destruction, and creation of new bonds in a new local particle equilibrium state.  Therefore, a physically-based continuum theory for non-stationary irreversible deformation should  be based on the micromechanics of these basic subprocesses. In this paper, we assume, however, that the characteristic times for the creation and destruction of bonds are negligibly small if compared with time $\tau$; therefore, these two subprocesses are eliminated from consideration.

Further to the physical model presented above, in order to characterize the degree of the deformation irreversibility or fluidity, we use the time $\tau$ of the ``settled life'' of particles (after Frenkel~\cite{Frenkel1955}), \textit{i.e.}, the time during which all bonds of a given particle with its neighbors are conserved; they can stretch but are not destroyed. The introduction of time $\tau$ allows the clear classification of all types of material responses from the same basis.  In fact, a pure elastic response corresponds to $\tau=\infty$ because all particle bonds are conserved,  thus the particles are not rearranged. If particle rearrangements occur, this automatically implies that $\tau<\infty$, \textit{i.e.}, $\tau$ becomes finite regardless of  whether  the nature of a material is  a viscous fluid or elastoplastic solid. Finally, the idealized limiting case $\tau=0$ corresponds to inviscid flows, \textit{i.e.}, ideal fluids or ideal plasticity in the case of solids, because bonds do not exist; their lifetime equals to zero. In other words, in order to characterize the degree of irreversibility of deformation, \textit{i.e.}, fluidity, we propose a temporal characteristic $\tau$ with a clear physical meaning that allows the comprehension of continuum dynamics of solids and fluids under a unified viewpoint.

Unlike the viscosity coefficient, the particle settled life (PSL) time represents an objective material characteristic that can be observed directly, at least potentially, in experiments without the introduction of any subjective notions invented by the observer, such as stresses. In addition, because of its universality or material-independent character, it covers  continuously the full spectrum of material responses, from inviscid flows ($\tau=0$) to pure elastic responses~($\tau=\infty$). 

The second critical material  parameter in our theory is connected with the most obvious difference between solid-like and fluid-like behavior, \textit{i.e.}, the ability of solids and inability of fluids to retain static shear loadings. In what follows, we explain that such a difference has a quantitative rather than qualitative character if both solids and fluids are considered as a system of connected particles. In fact, under static loadings, bonds between particles of a solid are able to stretch to some degree and are not destroyed, which means that $\tau=\infty$, if the loadings are not sufficiently large. In that case, the solid is in an equilibrium static state. If bond stretching exceeds a certain limit, then bonds are destroyed (\textit{i.e.}, $\tau<\infty$) and particles rearrange. Thus, we introduce a material parameter $Y_0$, which is the bond stretch limit under static loadings. From the previous discussion, it is clear that $Y_0>0$ and $\tau=\tau(Y_0)$ for solids.

In fluids, with a good approximation, $Y_0=0$, which means simply that $\tau<\infty$ for any applied external loadings, and the situation when $\tau=\infty$ is  impossible. For example, if a fluid is treated as a system of particles with bonds,  where the bonds are the diffusion mass fluxes between neighboring particles, then such bonds do not manifest themselves, \textit{i.e.}, $Y_0=0$,  in a static state because there is no change in macroscopic motion state, and thus inertial forces equal to zero.

A typical flow model after external loading is applied to a viscous fluid as follows. Because viscous fluids are characterized by $0<\tau<\infty$, there is no flow, \textit{i.e.}, particle rearrangement, at time $t<\tau$. Therefore, at a given strain rate $\dot\varepsilon$, bonds are stretched elastically during the interval of time $t\in[0,\tau]$. At the time $t=\tau$, the bond stretching reaches a value $Y=Y(\dot\varepsilon)>Y_0=0$, called here the dynamic bond stretch limit, at which bonds are destroyed. In turn, this results in particle rearrangements. Thus, this leads to the key point of our theory, which is \textit{the internal resistance to flow (viscosity) of fluids is a result of elastic stretching of microscopic bonds between fluid particles, and the macroscopic flow is a result of bond destructions and rearrangements of particles.}

In summary, with regard to the PSL time $\tau$ and static bond stretch limit $Y_0$, fluids and solids can be ranged as follows:
\begin{itemize}
\setlength{\itemsep}{5pt}
\item $\tau = \infty$, elastic solids;

\item $\tau(Y_0)=\left\{\begin{array}{ll}
\tau=\infty,& {\rm if\ bond\ stretching} < Y_0\\
0<\tau<\infty, & {\rm if\ bond\ stretching} \geq Y_0
\end{array}\right.$, elastoplastic solids;

\item $\tau(Y_0)=\left\{\begin{array}{ll}
\tau=\infty,& {\rm if\ bond\ stretching} < Y_0\\
\tau=0, & {\rm if\ bond\ stretching} \geq Y_0
\end{array}\right.$, ideally plastic solids;

\item $0<\tau<\infty$, $Y_0=0$, viscous fluids (Newtonian and non-Newtonian);

\item $\tau=0$, $Y_0=0$, ideal fluids;
\end{itemize}
In general, time $\tau$ is a function not only of $Y_0$, but it also depends on a current flow state, or state variables, and should be defined with the help of non-equilibrium statistical physics, or kinetic theory, \textit{etc.} In the case of Newtonian fluids, we shall demonstrate in Sect.~\ref{sec:Numexper} that in order to fulfil Newton's viscous law, the particle settled life time $\tau$ should depend on the mass density $\rho$ and entropy $s$, and does not depend on the distortion $\AAA$. Plausibly, this is not true for more complex flows like turbulent flows, and $\tau$ is a function of $\rho$, $s$, and $\AAA$ as well as of some other, so far unknown, state parameters, \textit{e.g.}, see Remark~~\hyperref[rem:Slips]{1} in Sect.~\ref{sec:State_var}.
The non-Newtonian behavior of viscous fluids is beyond the scope of this study, but it is clear that such fluids also belong in the class characterized by $0<\tau<\infty$, $Y_0=0$, and the same theory that is described in the following is applicable if the function $\tau$ of the state variables is properly defined, \textit{e.g.}, see~\cite{Leonov1976,Leonov1987}.


\subsection{Navier-Stokes equations and hyperbolicity}

It is  difficult to overestimate the paramount role of the Navier-Stokes equations (NSEs) in fluid dynamics and it seems that there is a common belief that the NSEs are the single model for viscous flows in the framework of classical continuum theory.

On the other hand, as with any continuum models, the NSEs have limited applicability, and one of the main goals of the fluid dynamics community is to discover these limits beyond which an application of the NSEs are questionable. 

There is no doubt that turbulence is one of the challenging and still poorly understood problems in the dynamics of viscous fluids. However, nearly a century of extensive, and mainly unsuccessful, searching for  the ultimate Navier-Stokes-based turbulence model~\cite{Tsinober2009} may be considered as evidence that turbulent flows might be one of the limits of applicability of the NSEs. Clearly, such a lengthy duration may signify not only the defects in the concept of the viscosity coefficient itself, but the intrinsic extreme difficulties of the turbulence problem.

Regardless of the case, it is clear that the widespread and unconsidered use of the NSEs may be dangerous because of the following drawbacks, which incidentally, can be one of the reasons suppressing further progress in the understanding and modeling of turbulent flows:

\begin{itemize}
\item \textit{Stress tensor.} The Navier-Stokes viscous stress tensor was derived from observations of steady and structureless/homogeneous flows. Successive applications of the NSEs to numerous practically important situations  also suggests that Newton's viscous law is a good approximation to non-steady near-equilibrium laminar flows. However, it becomes clear that the use of Navier-Stokes-like models for essentially unsteady and non-equilibrium flows, such as strongly sheared and time-dependent turbulent flows, is problematic~\cite{YakhotExa}.
\end{itemize}
The root of the problem is that the derivation of the NSEs' stress tensor is based on the strategy where the material responses observed experimentally are straightforwardly mimicked in a model instead of simulating the intrinsic reasons leading to the observable macroscopic response. Clearly, such a mimic strategy is applicable provided the circumstances are simple and we are able to capture the key details of the material response, which unlikely to be the case for turbulence. See the following section for a further discussion.
\begin{itemize}
\item \textit{Causality.} Solutions to the NSEs are of a diffusive nature and, hence, do not satisfy causality\footnote{Although linear parabolic diffusion theories based on Fourier, Fick, or Newton's laws predict that disturbances can propagate at infinite velocity, nonlinear parabolic equations,
in some instances, lead to finite velocity of propagation, \textit{e.g.}, see the discussion in~\cite{Gomez2010} and references therein.}, \textit{i.e.}, the NSEs admit infinite velocity of disturbance propagation, \textit{e.g.}, see~\cite{Jou2010,Muller1998,Gomez2010,Jordan2000,Romatschke2010,HuovinenMolnar2009}, while by its very nature, a flow of a fluid is a pure hyperbolic process; all disturbances propagate at finite velocities. Thus, to construct a pure hyperbolic model for viscous flows is of fundamental interest. Clearly, such a model will have several sound velocities: one for longitudinal and two for shear perturbations. 

\end{itemize}


Furthermore, it is important to emphasize that in the solid dynamics, it is a conventional point of view that a continuum model should be an obligatory hyperbolic. Not only because of the causality issue but also because of that the hyperbolicity notion and the local well-posedness of the boundary-initial value problem\footnote{With sufficiently smooth initial data and dissipative boundary conditions.} for a system of PDEs are convertible terms~\cite{GodUrMathPhys,LeFloch1988,Dafermos2005}. 
With regard to this, one questions what is so special in the physics of viscous fluids that makes them intrinsically parabolic. As we shall demonstrate in the following, there are no special characteristics and it is indeed possible to develop a pure hyperbolic theory for viscous flows.

It is worth noting that several approaches to construct a hyperbolic alternative to the parabolic NSEs are known, \textit{e.g.},~\cite{IsraelStewart1979,Jou2010,Muller1998,Romatschke2010,Morro,Geroch1995,Gomez2007,Nishikawa2010,Toro2014}. The central idea of these approaches is an introduction of the fluxes as additional independent state variables, and the search for general transport laws taking the form of evolution equations for these fluxes. All of these theories are essentially variations of the phenomenological idea of Maxwell~\cite{Maxwell1867} to model the viscoelastic response of fluids, and sometimes referred to as \textit{Extended Irreversible Thermodynamic} (EIT)~\cite{Jou2010}. Thus if we ignore the tensorial structure of the governing equations from the just cited papers, the equations have the general form 
\begin{equation}\label{eq:Maxwell_idea}
\dot{X}=-\frac{1}{\lambda}(X-X^{NS}) 
\end{equation}
for a dissipative quantity $X$, where $X^{NS}$ is the value of $X$ obtained in the framework of Navier-Stokes theory. 
In other words, in a Maxwell-type theory, a hyperbolic relaxation model (\ref{eq:Maxwell_idea}) is constructed as an first-order extension of the second-order NSEs in such a way that the NSEs is obtained as the relaxation limit of (\ref{eq:Maxwell_idea}) as a relaxation parameter $\lambda$ goes to zero. Note that (\ref{eq:Maxwell_idea})  is also referred to as Maxwell-Cattaneo law, because the same mathematical idea was applied by Cattaneo~\cite{Cattaneo1948} to develop a hyperbolic theory of heat conduction.

Even though there is a solid microphysics background as the kinetic theory of gases in some of the hyperbolic approaches, \textit{e.g.},~\cite{Jou2010,Romatschke2010}, and they are in good agreement with many experimental data, see textbooks~\cite{Jou2010,Muller1998}, the resulting macroscopic models are rather phenomenological, because their derivations are based on the series expansion of the distribution function with respect to an abstract infinite hierarchy of moments, in which the physical meanings of the introduced state variables, moments, are mostly ignored. For example, in the mentioned theories, the stress tensor is used as the state variable $X$, which, by its definition, is defined on specifically oriented hypersurfaces. The use of such state variables is in a contradiction with the central notion of the continuum physics, material particle (infinitesimal control volume), whose macroscopic properties, \textit{i.e.}, state parameters, should be defined by only an averaging over the volume of observable microscopic characteristics. The stress tensor neither is observed\footnote{Do not be confused with the measurable quantities, \textit{e.g.}, we can measure stresses, but they clearly do not exist, \textit{i.e.}, they are the result of an interpretation of the microscopic reality. Of course, the stress tensor arises in our theory as the momentum flux, but not as an independent state variable.} in reality, \textit{i.e.}, it does not have a microscopic counterpart, nor is defined as an volume averaged quantity (it is defined on specifically oriented hypersurfaces).


One more drawback, which is a consequence of the use of the state variables living on hypersurfaces, is that such models lack the objectivity property, \textit{i.e.}, frame invariance.  To treat such a discrepancy, an  objective stress rate $\dot{X}$ is usually introduced into (\ref{eq:Maxwell_idea}), which in turn can be defined in many ways resulting in different solutions to the model, incompatibility with the second law of thermodynamics, and even in loss of the well-posedness property of the initial value problem, see Sect.~\ref{sec:Maxwell} for the references. 

Finally, the models of type (\ref{eq:Maxwell_idea}) have unlimited growth of the shear sound velocity as the relaxation parameter $\lambda$ tends to zero (Newtonian limit). Meanwhile, there are no physical evidences supporting this point.  See a discussion in Sect.~\ref{sec:Maxwell} for more details and references. In this section, we only stress that neither the NSEs nor the concept of the Newtonian viscosity coefficient are used in any forms (\textit{e.g.}, like (\ref{eq:Maxwell_idea})) in the proposed theory, \textit{i.e.}, we do not modify the classical parabolic theory or treat its defects, but instead we propose an alternative intrinsically hyperbolic approach, which is based on the physical model described in the previous section, is free from the mentioned shortcomings, and involves less phenomenological constituents than the preceding approaches.



\subsection{\label{sec:resp_micro}Response modeling vs. microstructure modeling}

\paragraph{Response modeling approach.}

As mentioned previously, the derivation of the NSEs and NSE-like models in general is based on the imitation strategy when a mathematical model mimics the straightforwardly observed macroscopic response of a material in experiments. Such an approach provides quick results in situations closed to the conditions of experiments. However, extrapolation/prediction beyond the framework of these circumstances by means of variation of the constitutive law parameters becomes questionable when flow characteristics are changed non-linearly, \textit{e.g.}, laminar-turbulent transition. In general, such a mimic strategy should imply the analysis of all qualitatively different flows by experiments in order to develop a model that will be flow-independent, geometry independent, \textit{etc}. Therefore, there is no guarantee that the mathematical submodels derived for different flow regimes will continuously transform from one into another. In practice, this results in an excessively large number of models for a single material. 

We shall refer to such a mimic strategy as the response modeling (RM) approach. The RM approach is inherently an observer-dependent strategy, \textit{i.e.}, subjective. For example, typical state variables in a model developed in the framework of the RM approach are either the pressure $p$, temperature $T$, or total stresses $\TT$, \textit{etc}. 
However, it is clear that these entities do not exist in reality; they were invented by humans according to our inherent way of perception of the macroscopic reality by means of some interpretative procedures of microscopic processes, not to be confused with an averaging.

The classical representatives of the response modeling approach are the Maxwell, Oldroyd-B (\textit{e.g.},~\cite{Joseph1990}), Bingham, Hershel-Buckley~(\textit{e.g.}, see \cite{Frigaard2014}), Navier-Stokes, and hypoelastic Wilkins models~\cite{Wilkins1963,Rom1974,TrangColella1991,GavrFavr2008},  \textit{etc}. See also Sect.~\ref{sec:Maxwell} for a discussion on the relation between the Maxwell-type models and the model discussed in this paper.

\paragraph{Microstructure modeling approach.}

Intuitively, we understand that nature does not operate in terms of subjective state parameters and that continuum laws of physics should be written in terms of state parameters only, which are in principle, observable, without of engaging of any interpretative procedures. 
Thus, an alternative approach to the RM is not to straightforwardly mimic the macroscopic response of a material, but instead: \texttt{(i)} to comprehend what essential microscopic processes lead to such a macroscopic response; \texttt{(ii)} to determine objective measures, or state parameters, for these processes; and \texttt{(iii)}  to discover eventually the macroscopic evolution laws, which are usually hidden from  direct observation, for the state parameters defined in the previous step. In order to do this, the experimental observations of macroscopic material responses are no longer sufficient and additional  information of a universal character is required, \textit{e.g.}, consistency with thermodynamics, invariance under transformations of the physical space (Galilean or Lorentz invariance), causality, correctness of the initial value problem, \textit{etc.}, in addition to knowledge about the physics of the microstructure. We shall refer to such an approach as the microstructure modeling~(MM) approach.

A good example where the RM approach is barely applicable is complex fluids, \textit{i.e.}, fluids for which a complex internal structure, \textit{e.g.}, stretchable long-chain molecules, influences essentially  the macroscopic dynamics. In such media, microscopic processes, such as internal friction, rotations, and elastic stretching of molecules, are responsible for a single quantity that is the overall stress tensor. Thus, the relation between the stress tensor, or observable macroscopic material response, and microscopic processes is not a one-to-one relation and therefore, ill-posed. An attempt to derive a constitutive relation for the stress tensor in such a situation for a sufficiently large range of flows might lead to enormous complexity of the model. 

Finally, it is interesting to note that, in continuum thermodynamics, these two types of state variables, \textit{i.e.}, objective and subjective, are thermodynamically dual quantities in the following sense. A typical vector of state variables in an RM-based model is $\pp=(\vv,\TT,T)$, while in a corresponding MM-based model, a typical vector is $\qq=(\vv,\FF,\ww_1,\ww_2,\ldots,\BB_1,\BB_2, \ldots,s)$, where $s$ is the specific entropy, $\ww_i$ and $\BB_i$ are vectors and tensors representing essential, for macroscopic dynamics, microscopic processes (\textit{e.g.}, see~\cite{PeshGrmRom2014}). In addition, in MM approaches an additional quantity is required, \textit{i.e.}, a thermodynamic potential $\scU=\scU(\qq)$ usually playing the role of the total energy. Thus, the vectors $\pp$ and $\qq$ are related as
\[\vv=\scU_\vv,\ \ \TT=\scU_\FF+\scU_{\ww_1}+\scU_{\ww_2}+\ldots +\scU_{\BB_1}+\scU_{\BB_2}+\ldots,\ \ T=\scU_s.\]
See also discussions in Sections 3.5 and 3.6 in~\cite{PeshGrmRom2014}, \cite{Rom1974} and concluding remarks in~\cite{Besseling1968}.

Well-known examples of continuum models related to the MM approach are the Euler and ideal magnetohydrodynamics equations. Moreover, the model discussed in the following is also representative of the MM approach. For more complex examples, see~\cite{GrmelaOttingerI,GrmelaOttingerII,Grmela2010,PeshGrmRom2014} and references therein. 

\subsection{Motivation}

The aim of this paper is to draw the attention of the modeling community to the fact that the continuum description of viscous Newtonian flows is not restricted solely by the NSEs, and it is possible to simulate viscous fluids with a more physically-based model still within the framework of classical continuum mechanics, and which is compatible with the fundamental observations, such as thermodynamics, causality, and mathematical regularity.

The basis of our approach is the elastoplastic model of hyperelastic type with relaxation terms describing plastic deformation in solids, which was derived by Godunov and Romenski in~\cite{GodRom1972,God1978,GodRom2003} in order to simulate irreversible deformation of metals under extreme loading.

We note that  similar theories to~\cite{GodRom1972,God1978,GodRom2003} were developed  independently in~\cite{Eckart1948,Besseling1968,Leonov1976,Rubin1987} (see also~\cite{Leonov1987,Rubin1993} and references therein). The common feature of these theories is the use of a local strain field like $\AAA$ to characterize the interaction between material particles, and the differences concern only the choice of an internal energy function and a dissipation mechanism dictated by applications under consideration, \textit{e.g.}, elastoplastic solids~\cite{GodRom1972,God1978,Rubin1987} or polymeric liquids~\cite{Leonov1976}.    Eckart~\cite{Eckart1948} appears to have been the first to formulate a theory in terms of only an elastic deformation tensor, which can be used to characterize irreversible deformation in isotropic solids. However, he did not recognize that the internal resistance to flow was already presented in his theory, and the Newtonian viscosity was added. Besseling~\cite{Besseling1968} then resolved this misunderstanding and generalized this approach for anisotropic solids. He also showed that the viscous Newtonian flows can be described in the framework of his model.

This paper differs from the previously cited works in the following manner. In general, our goal is a three-dimensional finite-volume numerical implementation of the discussed hyperbolic model for viscous Newtonian flows and its application to the problems where the use of the NSEs might be questionable. Here, only the preliminary work with an emphasis on clear physical interpretation of the model constituents, hyperbolic nature of the model, its fully thermodynamic consistency, and some numerical aspects is reported. In addition, our study is based on the theory of thermodynamically compatible systems of hyperbolic conservation laws~\cite{God1961,GodRom2003,Rom1998,Rom2001,PeshGrmRom2014}. In this theory, the non-dissipative part of the time evolution is generated with the help of only one thermodynamic potential, which is assumed to be a convex function of the state variables. Here, we also show that the dissipative part of the time evolution obeys the second law of thermodynamics and can be generated by the same potential. Therefore, in this study only one thermodynamic potential is used, \textit{i.e.}, no an additional dissipative potential is required.

\section{\label{sec:Model}The model}

\subsection{\label{sec:State_var}State variables}
Let $\xx=(x_1,x_2,x_3)^\mathsf{T}$ denotes the laboratory Cartesian coordinate system. In order to characterize the macroscopic properties of a material particle located at a point $\xx$ at a time $t$, we use the following vector of state variables
\begin{equation}\label{eq:state_var}
(\rho\vv,\AAA,\rho,\rho s),
\end{equation}
where $\rho$ is the mass density of the medium, $\rho \vv$ is the momentum, $\vv$ is the velocity vector relative to the coordinate system $\xx$, $\rho s$ is the entropy density, $s$ is the specific entropy, and a newly adopted variable $\AAA$, non-symmetric $3\times3$-matrix, is the so-called elastic distortion~\cite{GodRom2003}. The field $\AAA$ has a local meaning, \textit{i.e.}, it describes the deformation and rotation of the material particles and, in general, is not related to the overall (global) deformation of the continuum. The locality of $\AAA$ also means that the fields $\vv$ and $\AAA$ are incompatible, \textit{i.e.}, knowing only field $\vv$ it is impossible to recover field $\AAA$ from its initial conditions if particle rearrangements occur. Note that if there are no particle rearrangements, and therefore there is no flow, then fields $\vv$ and $\AAA$ are compatible. In this case, $\AAA$ gives the global deformation of the continuum.


From the physical view point, the particle rearrangements imply the presence of microscopic slips of one material particle, in fact small volume, (or a cluster of particles) relative to another particle (cluster of particles). Intuitively, it is clear that the shear mechanical fluctuations cannot propagate across the slip planes, and thus the distortions of the particles located along the both sides of a slip plane become incompatible. That is to say, a part of strains dissipates in such slips. A possible mechanism of strain dissipation is described in Sect.~\ref{sec:strain_diss}. 

\paragraph{\label{rem:Slips} Remark \arabic{mypar}.}\stepcounter{mypar} \textit{
It may happen that a more microscopic description than (\ref{eq:state_var}) is required, \textit{e.g.}, for modeling of turbulent flows. Such a theory should be able to predict the material particle trajectories more accurately. This can be done if only more microscopic details  regarding the mechanism of particle rearrangements are taken into account. Thus, the particle rearrangements causing the macroscopic flows in a system of connected particles are known to occur in collectively organized manner~\cite{langerAthermal2007}, \textit{i.e.}, the bonds of a given particle with its neighbors are destroyed not simultaneously, and hence the particles are organised into clusters while they rearrange. In other words, the dynamics and interaction of the slips (called flow defects) of clusters of particles relative to other clusters of particles should be taken into account in a more microscopic continuum model, \textit{e.g.},~see system~(59) in \cite{PeshGrmRom2014} for a possible set of governing equations for such a theory.
}

Finally, since the material particles, by its definition, have a constant mass, then it is also assumed that the distortion $\AAA$ obeys the constraint 
\begin{equation}\label{eq:detA}
\rho=\rho_0\det\AAA,
\end{equation}
where $\rho_0$ is the reference (at the initial moment of time) mass density.

\subsection{\label{sec:time_evol}Time evolution}

Let  $\scE$ be the specific total energy of the system, which is a function of state variables (\ref{eq:state_var}), \textit{i.e.}, $\scE=\scE(\rho\vv,\AAA,\rho,\rho s)$. In addition, let $\scE_{\AAA}=[\scE_{A_{ij}}]$ denotes the matrix for which entries are the partial derivatives, $\partial \scE/\partial A_{ik}$, $\delta_{ik}$ denotes the Kronecker symbol, and $\tau=\tau(\AAA,\rho,s,Y_0)$ represents the characteristic time of strain dissipation, which  is a continuum interpretation of the particle settled life time introduced in Sect.~\ref{Intro}, $Y_0$ is the static stretch limit (also, see Sect.~\ref{Intro}). Therefore, in the Eulerian framework, the model is (\textit{e.g.}, see~\cite{God1978,GodRom2003})
\begin{subequations}\label{eq:model_eul}
\begin{align}
&\displaystyle\frac{\partial \rho v_i}{\partial t}+\frac{\partial (\rho v_i v_k+\rho^2\scE_\rho\delta_{ik}+\rho A_{mi}\scE_{A_{mk}})}{\partial x_k}=0, \label{eq:model_eul_a}\\[2mm]
&\displaystyle\frac{\partial A_{i k}}{\partial t}+\frac{\partial A_{im} v_m}{\partial x_k}=-v_j\left(\frac{\partial A_{ik}}{\partial x_j}-\frac{\partial A_{ij}}{\partial x_k}\right)-\dfrac{\Psi_{ik}}{\tau},\label{eq:model_eul_b}\\[2mm]
& \frac{\partial \rho}{\partial t}+\frac{\partial \rho v_k}{\partial x_k}=0,\label{eq:model_eul_cont}\\[2mm]
&\displaystyle\frac{\partial \rho s}{\partial t}+\frac{\partial \rho s v_k}{\partial x_k}=\rho\dfrac{\scE_{A_{ij}}}{\scE_s}\dfrac{\Psi_{ij}}{\tau}. \label{eq:model_eul_d}
\end{align}
\end{subequations}
The first equation represents the momentum conservation. The sum of the second and third terms in the momentum flux,
\begin{equation}\label{eq:stress_total}
\TT=[T_{ik}]=[-\rho^2\scE_\rho\delta_{ik}-\rho A_{mi}\scE_{A_{mk}}]=-\rho^2\scE_\rho\II - \rho\AAA^\mathsf{T}\scE_\AAA=\TT^\mathsf{T}
\end{equation}
is referred to as the symmetric total Cauchy stress tensor, $\rho^2\scE_\rho=p$ is the pressure, and $\II$ is the identity tensor.
Equation~(\ref{eq:model_eul_b}) is the time evolution equation for the elastic distortion $\AAA$, the last term on the right-hand side of (\ref{eq:model_eul_b}) represents the strain dissipation mechanism due to the particle rearrangements. Equations~(\ref{eq:model_eul_cont}) and (\ref{eq:model_eul_d}) are the continuity equation and the time evolution for the entropy density, respectively.

On the solution of system (\ref{eq:model_eul}), an additional conservation law is satisfied, which is the conservation of the total energy density $\rho\scE$ or the first law of thermodynamics:
\begin{equation}\label{eq:energy_cons}
\dfrac{\partial \rho \scE}{\partial t}+\dfrac{\partial }{\partial x_k}\left(v_k\rho \scE+\rho v_n(\rho\scE_\rho\delta_{nk} + A_{mn}\scE_{A_{mk}})\right)=0,
\end{equation}
It can be derived from system (\ref{eq:model_eul}) by  summing all Eqs.~(\ref{eq:model_eul}) multiplied by the respective multiplicative factors~\cite{GodRom1998,GodRom2003,Rom1998,Rom2001,PeshGrmRom2014}:
\[(\rho\scE)_{\rho v_i},\  (\rho\scE)_{A_{ik}},\ \scE - v_i\scE_{v_i}-V\scE_V-s\scE_s,\ (\rho\scE)_{\rho s},\]
where $V=\rho^{-1}$ is the specific volume. This also explains the origins of the source term in the entropy equation, \textit{i.e.}, if this source term is chosen in a different way, then we cannot guarantee the absence of a source term in the energy conservation law, which clearly would violate the first law of thermodynamics. In addition, we shall show in the following that the source term in the entropy equation is non-negative for an appropriate choice of the strain dissipation function $\PSI=[\Psi_{ik}]$, which makes the model fully thermodynamically consistent.

\paragraph{\textit{Remark \arabic{mypar}.}}\stepcounter{mypar} \textit{The mass density $\rho$ has been defined previously by $\rho=\rho_0\det\AAA$. If we multiply Eq.~(\ref{eq:model_eul_b}) by $\rho_{A_{ik}}$ (note that $\rho_\AAA=[\rho_{A_{ik}}]=\rho\AAA^{-\mathsf{T}}=\rho\EE^{\mathsf{T}}$) and sum them, we arrive at the mass conservation equation:
\begin{equation}\label{eq:continuity}
\frac{\partial \rho}{\partial t}+\frac{\partial \rho v_k}{\partial x_k}=0.
\end{equation}
This fact means that the mass conservation law is a consequence of (\ref{eq:model_eul_b}) and plays the role of the  involution constraint for the system (\ref{eq:model_eul}). However, it simplifies the situation, e.g., numerical implementation of the model, if we include the density $\rho$ in the set of state variables (\ref{eq:state_var}) for the reasons discussed in~\cite{MillerColella2001,BartonRom2010}.}

Applications of the model to the modeling of the elastoplastic deformation of solids can be found in~\cite{Pesh2010,BartonRom2010,FavrGavr2012,BartonRom2012,Barton2013}. A more sophisticated mechanism of irreversible deformation by means of the stress driven solid--fluid transition is described in~\cite{PeshGrmRom2014}. For the remainder of this paper, we discuss applications of the system (\ref{eq:model_eul}) to viscous Newtonian fluids.

\subsection{\label{sec:EOS}Equation of state for viscous fluids}

To complete the model, we specify the thermodynamic potential $\scE$, which plays a role of the equation of state or a closure relation in our continuum theory, and postulates the interaction rules for fields (\ref{eq:state_var}). In a proper continuum model such a potential should be derived with the help of microscopic theories as the non-equilibrium statistical physics, kinetic theory, \textit{etc.} So far, we can use only our intuitions and experience of construction of such potentials in the solid dynamics theory.

Let us first note that, as it is seen from the momentum equation, the total stress tensor $\TT=-\rho^2\scE_\rho\II-\rho\AAA^\mathsf{T}\scE_\AAA$ is generated by the total energy potential. Moreover, we shall show in the following that in order to guarantee the fulfilment of the second law of thermodynamics, the strain dissipation function $\PSI$ should be also generated with the help of~$\scE$. Thus, the specification of the total energy is a critical step in the model formulation. This is not surprising, because the equation of state should ``absorb'' the information about those essential microscopic processes, which occur below the resolution of the continuum scale and, in fact, govern the macroscopic dynamics. 

Traditionally, we assume the additive decomposition of the specific total energy into kinetic $\vv^2/2$ and internal energy $E(\AAA,\rho,s)$:
\begin{equation}\label{eq:energy_total}
\scE(\rho\vv,\AAA,\rho,\rho s) = \dfrac{\vv^2}{2} + E(\AAA,\rho,s).
\end{equation}
In turn, the following decomposition of the internal energy $E$ into the hydrodynamic part $E^h$ and the viscous part $E^v$  is assumed:
\begin{equation}\label{eq:energy_int}
E=E^h(\rho,s) + E^v(\AAA,\rho,s).
\end{equation}
For the hydrodynamic part $E^h$, any classical equations of state can be used, \textit{e.g.}, ideal gas, stiffened gas, \textit{etc.}, and the viscous part $E^v$ is considered in the following form:
\begin{equation}\label{eq:energy_visc}
E^v = \dfrac{c_s^2}{4}\left(\calI_2 - \dfrac{\calI_1^2}{3}\right)=\dfrac{c_s^2}{4}\hspace{1mm}{\rm tr}(({\rm dev}\,\GG)^\mathsf{T}{\rm dev}\,\GG),
\end{equation}
where
\[\begin{array}{c}
\calI_1={\rm tr}\,\GG,\ \ \calI_2={\rm tr}\,\GG^2,\ \  \GG=\AAA^\mathsf{T}\AAA,\\[3mm]
{\rm dev}\,\GG=\GG - \dfrac{{\rm tr}\,\GG}{3}\II,
\end{array}
\]
$\II$ is the identity tensor, and $c_s=c_s(\rho,s)$ is the shear sound velocity at rest, which is the material parameter characterizing the resistance of the particle bonds to shearing. 

It is useful to note that if $\AAA$ has a singular value decomposition $\AAA=\UU\boldsymbol{\calA}\BV^\mathsf{T}$ with $\UU^\mathsf{T}\UU=\BV^\mathsf{T}\BV=\II$ and $\boldsymbol{\calA}={\rm diag}(a_1,a_2,a_3)$ the diagonal matrix with the singular values, or principal stretches, $a_i>0$ on the diagonal, then $E^v$ can be written as
\begin{equation}\label{eq:energy_visc_sing}
E^v=\dfrac{c_s^2}{12}\left((a_1^2 - a_2^2)^2 + (a_2^2 - a_3^2)^2 + (a_3^2 - a_1^2)^2\right).
\end{equation}

The viscous energy $E^v$ makes a contribution to the overall internal energy $E$ only if shear deformations are present. In turn, shear deformations are absent  if and only if 
\begin{equation}\label{eq:shear_0}
\GG=\dfrac{{\rm tr\,}\GG}{3}\II\ \ \  {\rm or} \ \ \ a_1=a_2=a_3.
\end{equation}
In the incompressible case, $\det\AAA=a_1a_2a_3=1$, condition (\ref{eq:shear_0}) is equivalent to $a_1=a_2=a_3=1$,  thus $\GG=\AAA^\mathsf{T}\AAA=\II$, which essentially means that $\AAA$ is an orthogonal matrix.

We remark that $c_s$ is the velocity of propagation of pure elastic perturbations, \textit{i.e.}, perturbations that do not destroy bonds between fluid particles, see Sect.~\ref{sec:Discuss}, in the fluid at rest but not the velocity of propagation of observable diffusive shear perturbations. The elastic processes of microscopic bond stretching are weak, and besides are hidden, or masked, by the strong dissipative process of particle rearrangements, which makes the experimental measurements of the sound velocity $c_s$ a problematic procedure. However, it can be estimated theoretically using methods of non-equilibrium statistical physics. In addition, we shall discuss in Sect.~\ref{sec:Numexper} a way that $c_s$ can be calculated for a given value of the particle settled time $\tau$.

\paragraph{\textit{Remark \arabic{mypar}.}}\stepcounter{mypar}\label{rem:two_sound_vel} \textit{Finally, note that if a fluid flows, then there are two local shear sound velocities that become different from $c_s$ due to the factor $\calI_2-\calI_1/3$ in (\ref{eq:energy_visc}). Their values depend on the intensity of the shear flow (flow anisotropy), and can be computed as the characteristic velocities, or eigenvalues, of a quasilinear form of the system (\ref{eq:model_eul}); a standard procedure in the theory of hyperbolic PDEs~\cite{KulikPogorSemen}.}

\subsection{\label{remark1}Stress tensor}

The Cauchy stress tensor $\TT$ for the system (\ref{eq:model_eul}) is given by formula~(\ref{eq:stress_total}). 
The details of its derivation can be found in~\cite{God1978,GodRom1998,GodRom2003,PeshGrmRom2014}. Here, we only emphasize that the form of the Cauchy stress tensor is dictated by the requirement of compatibility with the first law of thermodynamics, \textit{i.e.}, energy conservation law (\ref{eq:energy_cons}) (\textit{e.g.}, see \cite{PeshGrmRom2014}). 

Taking into account (\ref{eq:energy_total}) and (\ref{eq:energy_int}), stress tensor (\ref{eq:stress_total}) can be  rewritten as
\begin{equation}\label{eq:stress_total2}
\TT=-p \II-\rho\AAA^\mathsf{T}E^v_{\AAA}:=\TT^h+\TT^v,
\end{equation}
where $\TT^h(\rho,s)=-p \II$ and $\TT^v(\AAA,\rho,s)=-\rho\AAA^\mathsf{T}E^v_{\AAA}$ are the hydrostatic and viscous parts of the total stress tensor, respectively. For a particular choice of $E^v$ given by  (\ref{eq:energy_visc}), the viscous stress tensor $\TT^v$ is
\begin{equation}\label{eq:Tvisc}
\TT^v=-\rho\, c_s^2\GG\left(\GG - \dfrac{\calI_1}{3}\II\right)=-\rho\, c_s^2\GG\left({\rm dev}\,\GG\right).
\end{equation}
As is seen from the definition of $\TT^v$, the viscous stresses are absent if $E^v_{\AAA}=\boldsymbol{0}$ or if~(\ref{eq:shear_0}) holds.

\paragraph{\textit{Remark \arabic{mypar}.}}\stepcounter{mypar} \textit{In general, the pressure $p=\rho^2\scE_\rho=\rho^2 E^h_\rho+\rho^2 E^v_\rho$ does not coincide with the classically defined isotropic pressure $p^{classic}=\rho^2 E^h_\rho$. For example, these two pressures will be different for strongly sheared flows due to the term $\rho^2 E^v_\rho=p-p^{classic}$. Clearly, the longitudinal sound velocity defined in the framework of the classical theory of viscous fluids $c_l^{classic}=\sqrt{\frac{\partial p^{classic}}{\partial\rho}}$ and the longitudinal sound velocity $c_l=\sqrt{\partial p/\partial\rho}$ for model (\ref{eq:model_eul}), (\ref{eq:energy_int}), in general, are also different.
}

\paragraph{\textit{Remark \arabic{mypar}.}}\stepcounter{mypar} \textit{In contrast to the NSEs, the hydrostatic $\TT^h$ and the viscous $\TT^v$ stresses are generated in a unified manner, \textit{i.e.}, as the partial derivatives of the total energy $\scE$ with respect to the state variables, the so-called thermodynamic forces. This is a standard feature of models developed in the framework of the MM ideology, in particular in the framework of the thermodynamically consistent theory of conservation laws (see~\cite{Besseling1968,Rom1998,Rom2001,GodRom2003,PeshGrmRom2014} and references therein) or in the GENERIC approach~\cite{GrmelaOttingerI,GrmelaOttingerII,GrmelaAfif2002,Grmela2010}.}

\subsection{\label{sec:strain_diss}Strain dissipation}

Consider the pure dissipative part of the overall time evolution (\ref{eq:model_eul}), \textit{i.e.},
\begin{subequations}\label{eq:time_evol_diss}
\begin{align}
&\displaystyle\dfrac{\partial \AAA}{\partial t}=-\dfrac{\PSI}{\tau},\label{eq:time_evol_diss_a}\\[2mm]
&\displaystyle\frac{\partial \rho s}{\partial t}=\rho\dfrac{\scE_{A_{ij}}}{\scE_s}\dfrac{\Psi_{ij}}{\tau}. \label{eq:time_evol_diss_b}
\end{align}
\end{subequations}
We impose the following three conditions on the strain dissipation function $\PSI=[\Psi_{ij}]$~\cite{GodRom2003}:
\begin{itemize}
\item[(D1)]
\textit{Viscous stresses $\TT^v$ should relax during the strain dissipation process (\ref{eq:time_evol_diss_a}).} This condition originates from the well-known Maxwell conjecture~\cite{Maxwell1867,GodRom2003}. 
\end{itemize}
We note that the conventional material models such as, the Oldroyd-B model in rheology or the hypoelastic Wilkins model in elastoplasticity, are based directly on this property, because they use the stress tensor as a state variable. According to the philosophy of the MM approach (see the Sect.~\ref{Intro}), we avoid the use of such a subjective quantity as the stress tensor as a state parameter in our theory. Instead, we simulate the cause, \textit{i.e.}, particle rearrangement, which then leads to the observable macroscopic phenomenon as stress relaxation.

\begin{itemize}
\item[(D2)]
\textit{The mass density $\rho$ is not affected by strain dissipation.} This condition is a reflection of the natural assumption that microscopic particle rearrangement cannot generate macroscopic motion. Because of the equality $\rho=\rho_0\det\AAA$, this condition can be reformulated as 
\[\dfrac{\partial \det \AAA}{\partial t}\equiv 0\]
during the strain dissipation process (\ref{eq:time_evol_diss_a}).
\item[(D3)]
\textit{Entropy should not decrease during the dissipative time evolution (\ref{eq:time_evol_diss}),} \textit{i.e.},  the right-hand side of (\ref{eq:time_evol_diss_b}) should be non-negative:
\begin{equation*}\label{eq:2law}
\frac{\partial \rho s}{\partial t}=\rho\dfrac{\scE_{A_{ij}}}{\scE_s}\dfrac{\Psi_{ij}}{\tau}\geq 0.
\end{equation*}
\end{itemize}

In what follows, we show that if the strain dissipation function $\PSI$ is chosen to be proportional to $\scE_\AAA=E_\AAA=E^v_\AAA$ with a non-negative proportionality coefficient, then the three conditions are satisfied, which  emphasizes the important role of the thermodynamic potential~$\scE$.  We construct
\begin{equation}\label{eq:PSI}
\dfrac{\PSI}{\tau} = \dfrac{3}{c_s^2\tau\Delta}E^v_\AAA=\dfrac{3}{\tau\Delta}\AAA ({\rm dev}\,\GG),\ \ \ \Delta=\det\AAA>0.
\end{equation}

It is clear that condition (D3) is automatically satisfied if $\PSI$ is given by formula (\ref{eq:PSI}) because the right-hand side of (\ref{eq:time_evol_diss_b}) becomes a quadratic form with positive coefficients. By comparing (\ref{eq:Tvisc}) and (\ref{eq:PSI}), it is obvious that condition (D1) is also satisfied. A proof of condition (D2) can be found in~\cite{PeshGrmRom2014}.

It still remains to specify the strain dissipation time $\tau=\tau(\AAA,\rho,s,Y_0)$. As it is mentioned in the introduction section, a proper derivation of the functional dependence  $\tau(\AAA,\rho,s,Y_0)$ should be based on the microscopic theories as the non-equilibrium statistical physics. Sometimes, it is, however, possible to obtain  interpolation formulas for $\tau$ from experimental data. For example, interpolation formulas for $\tau$ in metals can be found in \cite{GodDemchuk1974,GodKozin1974,GodDenisenko1975,Rubin1987,BartonRom2010,BartonRom2012,Barton2013}. As we have already discussed in Sect.~\ref{Intro}, the static stretching limit $Y_0=0$ in case of  fluids, which means that the strain dissipation mechanism (\ref{eq:time_evol_diss}) is turned on immediately as flow starts. Moreover, in Sect.~\ref{sec:Numexper}, we demonstrate that in order to fulfil Newton's viscous law, $\tau$ should be taken in the form $\tau=\tau(\rho,s)$, \textit{i.e.}, $\tau$ is constant for given values of $\rho$ and $s$ and does not depend on $\AAA$.

Finally, note that the equation~(\ref{eq:PSI}) provides that the distortion $\AAA$ and dissipation function $\PSI$ are coaxial. This means that the nine ordinary differential equations (\ref{eq:time_evol_diss_a}) equivalent to the following three differential equations:
\begin{equation}\label{eq:time_evol_sing}
\frac{\partial a_i}{\partial t}=-\frac{(2\,a_i^2 - a_m^2 - a_n^2)}{\tau a_m a_n},\ \ \ i\neq m\neq n\neq i,
\end{equation}
written in the terms of singular values $a_i$ of the elastic distortion $\AAA$.

\subsection{\label{sec:Hyperb}Hyperbolicity and stability of the equilibrium state}

Because of the Galilean invariance of the model~\cite{Ndanou2014}, the condition of hyperbolicity for system (\ref{eq:model_eul}) in  the three-dimensional case is equivalent to hyperbolicity in a one-dimensional case. In one-dimension, say $x_1$, system (\ref{eq:model_eul}) can be transformed into an equivalent quasilinear symmetric hyperbolic form if the total energy $\scE$ is a convex function with respect to the state variables $\rho\vv$, $\rho$, $\rho s$, and only the first column $A_{i1}$ of the elastic distortion $\AAA$~\cite{GodRom1998,Pesh2010}; or the second or third if $x_2$ or $x_3$-direction are chosen. In particular,  viscous energy (\ref{eq:energy_visc}) is a convex function of $A_{i1}$ if the following inequalities are satisfied~\cite{Pesh2010}; here, we use  Eq.~(\ref{eq:energy_visc_sing})
\begin{subequations}
\begin{align*}
&E_{a_1a_1}^v=\dfrac{1}{3}c_s^2\,(6\,a_1^2-a_2^2-a_3^2)>0,\\[2mm]
&\dfrac{E_{a_1}^v-E_{a_2}^v}{a_1-a_2}+\dfrac{E_{a_1}^v+E_{a_2}^v}{a_1+a_2}=\dfrac{2}{3}c_s^2\,(2\,a_1^2+2a_2^2-a_3^2)>0,\\[2mm]
&\dfrac{E_{a_3}^v-E_{a_1}^v}{a_3-a_1}+\dfrac{E_{a_3}^v+E_{a_1}^v}{a_3+a_1}=\dfrac{2}{3}c_s^2\,(2\,a_1^2+2a_3^2-a_2^2)>0,
\end{align*}
\end{subequations}
or if
\begin{equation}\label{eq:convexity_cond}
a_1^2>(a_2^2+a_3^2)/6\ \ {\rm and}\ \ a_3^2<2(a_1^2+a_2^2)\ \ {\rm and}\ \  a_2^2<2(a_1^2+a_3^2),
\end{equation}
which is sufficient for fluid dynamics applications because the singular values are close to one (see Sect.~\ref{sec:Numexper} for a typical order of $a_i-1$).
 Characteristic analysis of the model can be found in~\cite{BartonRom2010,FavrGavr2012}.

It is important to emphasize that convexity conditions (\ref{eq:convexity_cond}) also guarantee that the shear stress free equilibrium state characterized by the equality $\PSI=0$, or $E^v_{\AAA}=\boldsymbol{0}$, or  $a_1=a_2=a_3$ is a stable steady point for differential equation (\ref{eq:time_evol_diss_a}), \textit{i.e.}, the steady point is attractor. 

\subsection{\label{sec:Maxwell}Relation to Maxwell fluid}

Despite that fact that the development of the discussed hyperbolic theory in~\cite{GodRom1972,God1978,GodRom2003,GodDemchuk1974,GodKozin1974,GodDenisenko1975} was inspired by the Maxwell material model for viscoelastic fluids, these two theories have little in common. 

For the discussion that follows, it is useful to recall that in the Maxwell-type models the stress tensor plays the role of a state variable. This fact is automatically attributed such models to the RM approach (see Sect.~\ref{sec:resp_micro}). A typical time evolution for the shear stress tensor is
\begin{equation}\label{eq:Maxwell}
\dot{\boldsymbol{\Sigma}}=-\dfrac{1}{\lambda}(\boldsymbol{\Sigma}-2\eta\DD),
\end{equation}
where $\boldsymbol{\Sigma}$ is the shear stress tensor, $\dot{\boldsymbol{\Sigma}}$ is a frame invariant time derivative of $\boldsymbol{\Sigma}$, $\lambda$ is the so-called stress relaxation time, $\eta$ is the Newtonian viscosity coefficient, and $\DD$ is the symmetric part of the velocity gradient.

\paragraph{\textit{Stress relaxation vs. strain dissipation.}} As it is seen in (\ref{eq:Maxwell}), Maxwell type-models are primarily based on the concept of the viscosity coefficient, \textit{i.e.}, the NSEs, or viscous Newtonian flows, are the relaxation limit of a Maxwell-like model as $\lambda\rightarrow 0$.  In contrast, as we have already discussed in Sect.~\ref{Intro} and shall demonstrate numerically in the following section, the strain dissipation time, or PSL time, $\tau\neq0$ always for viscous flows in general and Newtonian flows in particular, and $\tau=0$ only in the case of inviscid fluids. Therefore, there cannot be a one-to-one relation between the times $\lambda$ and~$\tau$. 

The second discrepancy is the consequence of the first. The fact that the Maxwell-like models are based on the NSEs results in the unlimited growth of the velocity of shear perturbations as the stress relaxation time $\lambda$ tends to zero, which in turn makes such models inconsistent with experimental observations on wave propagation. Thus, for example, the shear sound velocity $c$ for the upper-convected Maxwell fluid is~\cite{Joseph1985}: $c=c(\lambda)=\sqrt{\eta/(\rho\, \lambda)}$. In particular, according to this formula, for some small value of $\lambda$, the shear sound velocity may become greater than the velocity of propagation of longitudinal sound waves, which is also physically meaningless. It should be noted that if however~(\ref{eq:Maxwell}) is not postulated phenomenologically but derived from the kinetic theory of gases~\cite{grad1949kinetic,Jou2010}, this paradox is eliminated. Nevertheless, this approach suffers from that the derived macroscopic theory does not satisfy the principle of material frame-indifference (objectivity), \textit{i.e.}, the time derivative $\dot{\boldsymbol{\Sigma}}$ is not a frame invariant time derivative.

In contrast, in the discussed hyperbolic theory, because the dissipative source term $\PSI/\tau$ does not involve any spatial derivatives $\partial/\partial x_k$,  it is clear then that the shear sound velocities (see Remark~\hyperref[rem:two_sound_vel]{3} in Sec.~\ref{sec:EOS}) do not depend on~$\tau$ and therefore they are always finite even though the limit of $\tau\rightarrow 0$ is considered, \textit{e.g.}, see~\cite{GavrFavr2008,BartonRom2010,Pesh2010} for a characteristic analyses of~(\ref{eq:model_eul}) for some typical examples of internal energy~(\ref{eq:energy_int}). See also formulas (\ref{eq:tau_cs}) and (\ref{eq:eta_tau}) in the following section for the relation between parameter $c_s$ (shear sound velocity at rest, see~(\ref{eq:energy_visc})), Newtonian viscosity coefficient $\eta$, and the strain dissipation time $\tau$. 

\paragraph{\textit{Mathematical regularity and thermodynamics.}} Although equivalence between the second law of thermodynamics and criteria for well-posedness of the Cauchy problem has been demonstrated in~\cite{Rutkevich1972} for the upper-convected Maxwell fluids, it is proved in~\cite{Dupret1986} that this equivalence does not hold, in general, for other  objective Maxwell-type models. Moreover, it is known that some frame invariant time derivatives $\dot{\boldsymbol{\Sigma}}$ are not compatible with the second law of thermodynamics~\cite{trangenstein1991higher}.

The situation is different in the discussed hyperbolic theory. As mentioned in Sects.~\ref{sec:time_evol}, \ref{sec:strain_diss}, and \ref{sec:Hyperb}, although, without rigorous proof (the proof can be found in the cited papers, \textit{e.g.},~\cite{GodRom1998,Rom2001,PeshGrmRom2014}), a proper choice of the thermodynamic potential $\scE$ provides both properties simultaneously: complete compatibility with thermodynamics and mathematical regularity (hyperbolicity) of the model.

\paragraph{\label{sec:nonequilibr}Near-equilibrium and far-from-equilibrium flows.}
Model~(\ref{eq:model_eul}) belongs in the class of non-equilibrium models due to the dissipative source terms~(\ref{eq:time_evol_diss}), (\ref{eq:PSI}) of relaxation type. If the fluid particles are distorted, but the macroscopic flow is absent, then the fluid governed by only dissipative time evolution~(\ref{eq:time_evol_diss}), (\ref{eq:PSI}) falls into an equilibrium state characterized by the equality ${\rm dev}\,\GG=0$, which means that the fluid particles are not distorted in any direction.

Such an equilibrium state does not exist if macroscopic flow is present and $\tau>0$. Instead, the system now tends to reach a new equilibrium state characterized by ${\rm dev}\,\GG\neq 0$, which is the result of coupling of non-dissipative (\textit{i.e.}, (\ref{eq:model_eul}) with $\PSI\equiv 0$) and dissipative time evolutions~(\ref{eq:time_evol_diss}), (\ref{eq:PSI}). In contrast to~(\ref{eq:Maxwell}), this equilibrium state is not explicitly prescribed to the model. We shall demonstrate in Sect.~\ref{sec:Numexper} that Newton's viscous law is nevertheless fulfilled for such implicitly defined equilibrium states.

If macroscopic flow is present, but $\tau=0$, then the flow is equilibrium with ${\rm dev}\,\GG=0$, \textit{i.e.}, the fluid particles rearrange but they are not distorted. Recall that the equality $\tau=0$ means that  the particle bonds do not exist, and therefore the particles do not interact\footnote{The volume interactions between particles are still present, but the tangential interactions are missing because the particles bonds do not exist.}. In other words, the relaxation limit, \textit{i.e.}, as $\tau\rightarrow 0$, of the discussed hyperbolic model is the Euler equations (inviscid fluid), which are also hyperbolic. Recall that the formal relaxation limit of~(\ref{eq:Maxwell}), \textit{i.e.}, as $\lambda\rightarrow 0$, is the Navier-Stokes equations\footnote{If however (\ref{eq:Maxwell}) is derived from the kinetic theory of gases then $\eta\rightarrow 0$ as $\lambda\rightarrow 0$, and the relaxation limit of (\ref{eq:Maxwell}) is also the Euler equations, \textit{e.g.}, see~\cite{Jou2010}.}. Due to this fact, mathematical problem~(\ref{eq:Maxwell}) is predominantly parabolic for near-equilibrium flows (\textit{i.e.}, Newtonian flows), while the developed theory is hyperbolic\footnote{See the following section for a criteria of hyperbolicity.} regardless of whether the flow is near equilibrium or not.


Finally, recall that system~(\ref{eq:model_eul}) was originally derived for modeling of non-linear elastoplastic deformations in solids~\cite{God1978,GodRom2003}. This means that there are no theoretical restrictions on the values of the distortion $\AAA$. Hence, the model can be potentially applied for simulation of far-from-equilibrium flows provided the viscous internal energy $E^v$ and the strain dissipation time $\tau$ are properly defined. Thus, note that formula~(\ref{eq:energy_visc})  is just a particular realization for $E^v$ found to be sufficient to demonstrate the fulfilment of Newton's viscous law (see Sect.~\ref{sec:Numexper}), but it may happen that a more general (more non-linear) form of $E^v$ is required for strongly sheared non-equilibrium flows.

\paragraph{\textit{Physical interpretation.}} All elements of our continuum hyperbolic theory,  such  as elastic distortion of fluid particles and particle settled life time, are observable in principle. In contrast, the Maxwell-type models are representatives of the RM approach (see Sect.~\ref{Intro}) and hence, phenomenological.

\section{\label{sec:Numexper}Numerical examples}

The main goal of this section is to demonstrate that Newton's viscous law can be recovered with our hyperbolic theory. The second goal is to discover the functional dependence $\tau=\tau(\AAA,\rho,s)$ for which this can be done.

Consider a simple shear flow of a layer of a Newtonian fluid
\[\dot \varepsilon=\dfrac{\partial v_2}{\partial x_1},\]
where $\dot\varepsilon=\partial \varepsilon/\partial t$ is the rate of strain. Then, we rewrite Eq.~(\ref{eq:model_eul_b}) for elastic distortion $\AAA$ in the form
\begin{equation}\label{eq:A_noncons}
\dfrac{\partial A_{ij}}{\partial t}+v_k \dfrac{\partial A_{ij}}{\partial x_k} + A_{ik} \dfrac{\partial v_k}{\partial x_j}=-\dfrac{\Psi_{ij}}{\tau}.
\end{equation}
Given that $v_1=v_3\equiv0$ and $\partial/\partial x_2=\partial/\partial x_2\equiv0$,  the equations for the first column of $\AAA$ have the form
\begin{eqnarray}\label{eq:1_column}
\dfrac{\partial A_{i1}}{\partial t} + \dot{\varepsilon} A_{i2} = - \dfrac{\Psi_{i1}}{\tau},
\end{eqnarray}
and equations for the second and third columns are
\begin{eqnarray}\label{eq:23_column}
\dfrac{\partial A_{ij}}{\partial t} = - \dfrac{\Psi_{ij}}{\tau},\ \ i=1,2,3,\ \ j=2,3.
\end{eqnarray}

Finally,  we  solve  the  system (\ref{eq:1_column}) and (\ref{eq:23_column}) of nine ordinary differential equations with the rate of strain $\dot{\varepsilon}$ as a parameter. In what follows, formulas (\ref{eq:PSI}) and (\ref{eq:Tvisc}) for the strain dissipation function $\PSI$ and the viscous stress tensor $\TT^v$, respectively, are used. For our study, heat effects can be neglected, and because the flow is incompressible, the shear sound velocity $c_s$ is assumed to be constant. 

Since strain dissipation occurs in the direction of minimization of $\|{\rm dev}\,\GG\|$\footnote{In general, the directions of minimization of $\|{\rm dev}\AAA\|$ and $\|{\rm dev}\,\GG\|$ do not coincide.}, we can define the norm of ${\rm dev}\,\GG$ as the measure of the dynamical stretch limit $Y$ for a given fluid particle, \textit{i.e.},
\[Y=\|{\rm dev}\,\GG\|=\sqrt{{\rm tr}(({\rm dev}\,\GG)^\mathsf{T}{\rm dev}\,\GG)}.\]

For the numerical resolution of (\ref{eq:1_column}),~(\ref{eq:23_column}), the open source ODE solver LSODE~\cite{LSODE} and the solver ODE15S of the commercial software MATLAB~\cite{MATLAB} for stiff ODEs were used. Both solvers provided identical results.

\subsection{\label{sec:simp_shear}Simple shear flow}

As a representative of Newtonian fluids, we consider air with the viscosity $\eta=18.21\cdot 10^{-6}$~Pa$\cdot $s and reference density $\rho_0=1.2$~kg/m$^3$. Thus, we solve system (\ref{eq:1_column}), (\ref{eq:23_column}) with the initial condition $\AAA=\II$ and for different values of the rate of strain: $\dot{\varepsilon}=0$, $2$, $4$, $6$, $8$, $10$ s$^{-1}$.

\begin{figure}[t]
\includegraphics[trim = -10mm 180mm 0mm 10mm, clip, scale=0.7]{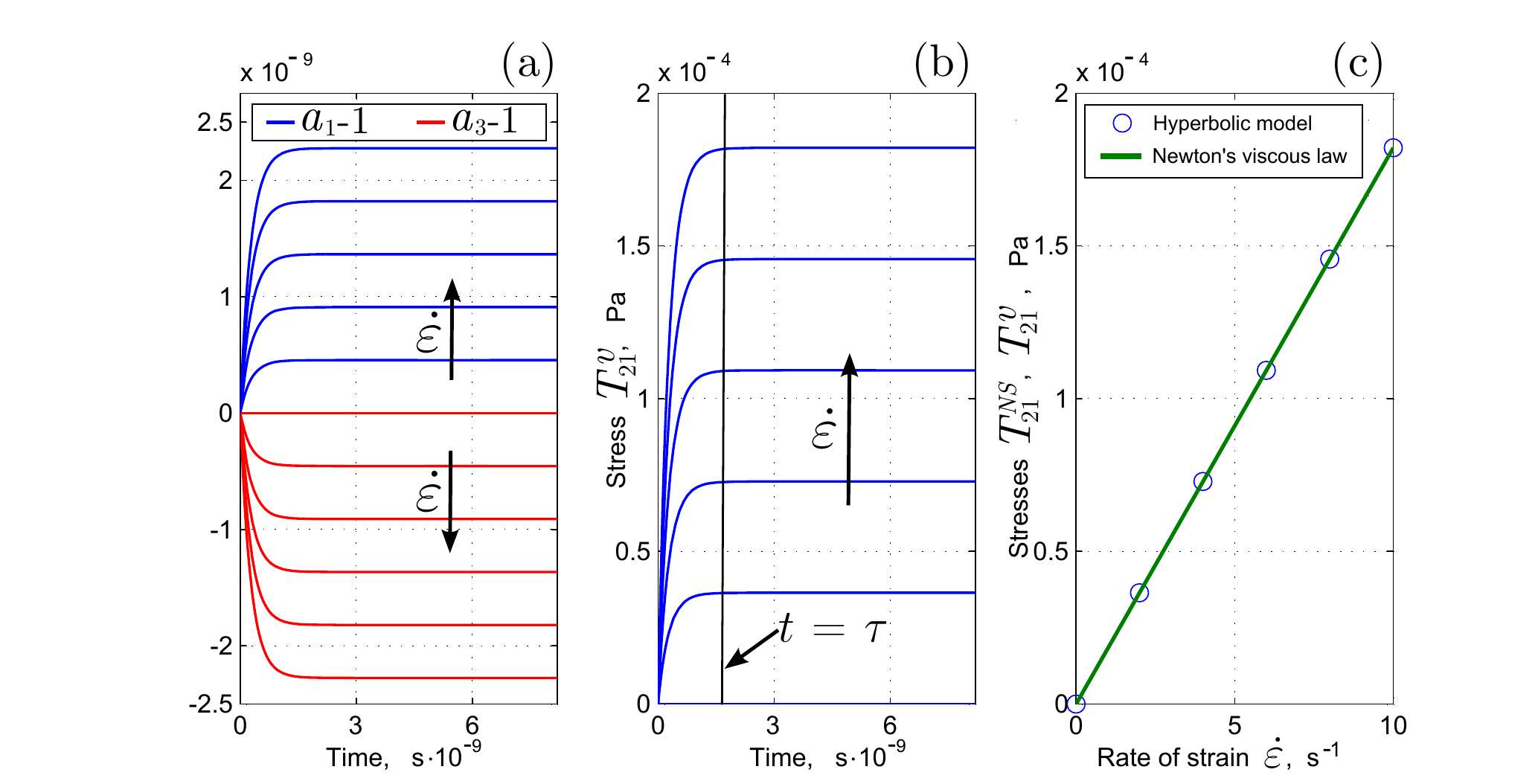}
\caption{\it \footnotesize Results of the numerical solution of system (\ref{eq:1_column}), (\ref{eq:23_column}) for the rate of strain $\dot{\varepsilon}=0$, $2$, $4$, $6$, $8$, $10$ s$^{-1}$ for air with the viscosity coefficient $\eta=18.21\cdot 10^{-6}$ Pa$\cdot $s: (a) time evolutions of the first and third singular values of matrix $\AAA$ (the second singular value equals identically to one in this test), (b) profiles of the shear stress $T_{21}$, (c) comparison of Newton's viscous law for air (green solid line) with results obtained with the hyperbolic model (blue circles, the steady-state values for $T_{21}$ extracted from plot~(b)).}\label{fig:fig_air}
\end{figure}

Our goal is to fit the viscous stress $T^v_{21}=T^v_{21}(\AAA)=T^v_{12}$ obtained as a solution to the hyperbolic model, \textit{i.e.} as a solution to system (\ref{eq:1_column}), (\ref{eq:23_column}),  to  Newton's viscous law 
\begin{equation}\label{eq:T_21}
T^{NS}_{21}=\eta\,\dot{\varepsilon}.
\end{equation}
The result of the fitting is presented in Fig.~\ref{fig:fig_air}(c). It appears that, in order to fit the law (\ref{eq:T_21}), the strain dissipation time can be assumed to be constant for a given density and temperature, at least for moderate values of rates of strains. In particular, for air $\tau=1.45\cdot10^{-9}$~s, provided the shear sound velocity\footnote{Recall that the longitudinal sound velocity of air is about $343$ m/s at $20$ $^{\rm o}$C.} is $c_s=250$~m/s. Results for other typical Newtonian fluids, such as water and a liquid with the viscosity coefficient close to a value typical for honey, are given in Table~\ref{tab:tab1}. 

In the following, we present the first observation in our numerical experiments.
\paragraph{\textit{Observation 1.}}\stepcounter{mypar} \textit{
Different values of the shear sound velocity provide different values of $\tau$, but our calculations show that the quantity $\tau c_s^2$ remains constant,
\begin{equation}\label{eq:tau_cs}
\tau c_s^2 =\nu_0,
\end{equation}
for a given Newtonian fluid. Clearly, this is because we are fitting one parametric law (\ref{eq:T_21}) with the two parameter model. The physical dimension of $\nu_0$ is m$^2$/s.
}

It may seem that any values of the shear sound velocity are appropriate provided that equality (\ref{eq:tau_cs}) is fulfilled. However, it is important to recall that $\tau$ cannot be arbitrary, since it is an objective quantity with a definite physical meaning, and thus can be estimated theoretically or in experiments. For example, water is a hardly compressible fluid, hence one may expect that the shear characteristics, \textit{e.g.}, shear sound velocity $c_s$, should be sufficiently smaller than the longitudinal characteristics, \textit{e.g.}, the longitudinal sound velocity $c_l$, say $c_s=c_l/10\approx1500/10=150$ m/s. In contrary, air is an easily compressible fluid, and hence one may expect that the air shear sound velocity is of order of longitudinal air sound velocity. For example, we have chosen $c_s=0.73\cdot c_l=250$ m/s for our numerical experiments in the case of air, see Table~\ref{tab:tab1}. 

The key feature of the model is that the microscopic elastic deformation, characterized by $\AAA$, of a fluid particle reaches rapidly (during the stage of elastic stretching $t<\tau$, see the vertical solid line on Fig.~\ref{fig:fig_air}(b)) toward a certain value corresponding to the dynamical stretch limit $Y$ and then remains constant while the observable macroscopic deformation of the fluid layer continues to increase proportionally with respect to the time of observation. Fig.~\ref{fig:fig_air}(a), (b) shows the time evolution of the first and third singular values $a_i$ ($a_2\equiv 1$ for this test case) of elastic distortion $\AAA$ (for the sake of convenient scaling, we plot $a_i-1$ instead of $a_i$) and the shear stress $T^v_{21}$, respectively for different values of $\dot{\varepsilon}$. 

This steady-state behavior of the singular values $a_i$ is in agreement with our interpretation of elastic distortion $\AAA$ as a parameter describing elastic stretching of microscopic bonds between fluid particles. Therefore, parts of bonds break because they have reached the dynamic stretch limit $Y(\dot{\varepsilon})$, while newly created bonds, for which the stretch is still below $Y$, continue to stretch. This process of bond stretching and destruction occurs permanently, but on average, it provides the current macroscopic steady-state value of the stress $T^v_{21}$. We also note that the steady-state values for the singular values $a_i$ remain quite close to 1, \textit{i.e.}, $a_i-1$ is of order $10^{-9}$ (see Fig.~\ref{fig:fig_air}(a)).

Fig.~\ref{fig:fig_air}(b) depicts the time evolutions of the shear stress $T^v_{21}$ for different values of~$\dot{\varepsilon}$. The steady-state values of $T^v_{21}$ are then plotted on Fig.~\ref{fig:fig_air}(c) with blue circles, and compared with $T^{NS}_{21}$ (green line).

Our second observation is:
\paragraph{\textit{Observation 2.}}\stepcounter{mypar} \textit{
For any Newtonian flows, the relation between the settled time $\tau$ and the viscosity coefficient $\eta$ is
\begin{equation}\label{eq:eta_tau}
\eta = \varrho_0\tau c_s^2=\varrho_0\nu_0,
\end{equation}
where the coefficient of proportionality $\varrho_0$ has the physical dimension of density, g/m$^3$. 
}

For particular choices (\ref{eq:energy_visc}) and (\ref{eq:PSI}) of the viscous internal energy $E^v$ and the strain dissipation function $\PSI$, $\varrho_0=\rho_0/6$.


\subsection{Arbitrary incompressible and compressible flows}

\begin{figure}[t]
\includegraphics[trim = -10mm 140mm 0mm 10mm, clip, scale=0.7]{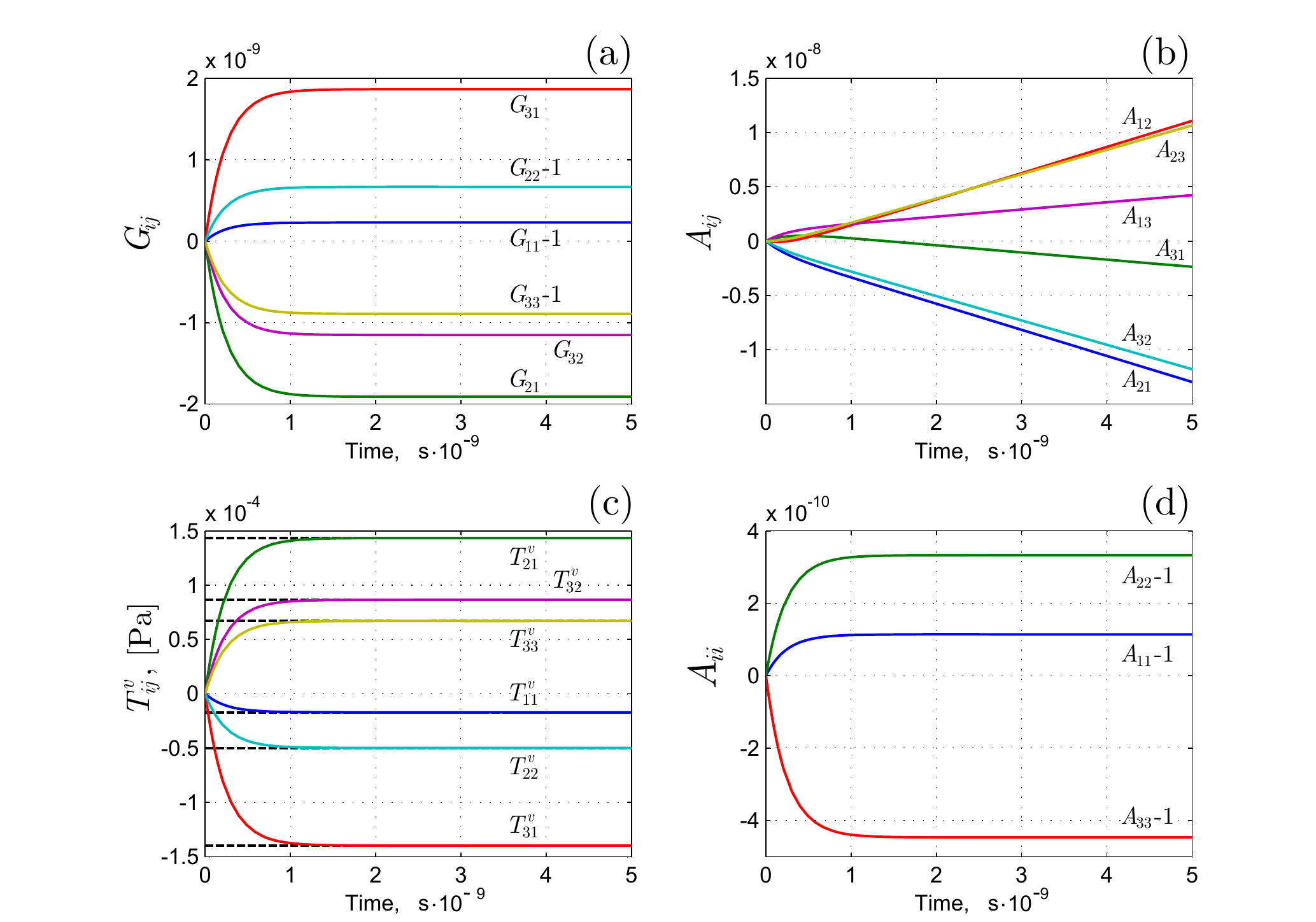}
\caption{\it \footnotesize Results of numerical solution of Eq.~(\ref{eq:A_noncons}) for an incompressible flow of air given by arbitrary chosen rate of strain tensor (\ref{eq:example}). Solution profiles on (a), (c), and (d) demonstrate a steady-state behavior at times $t>\tau$ while the off-diagonal entries $A_{ij}$, $i\neq j$ of the elastic distortion $\AAA$ continue to grow,  (b), which essentially means that the fluid particles rotate during they rearrange with neighbors. The black dashed lines in (c) are the components of the classical viscous stresses $\TT^{NS}$.}\label{fig:fig2} 
\end{figure}

We first examine the model in the case of incompressible flows. In this example, the rate of strain tensor
\begin{equation}\label{eq:example}
\dot{\boldsymbol{\varepsilon}}\equiv\left[\dfrac{\partial v_i}{\partial x_j}\right]=\left[
\begin{array}{rrr}
-0.47	&	 1.53	&	-4.50	 \\
 6.34	&	-1.37	&	 0.13	 \\
-3.18	&	 4.62	&	 1.84	
\end{array}
\right]\ {\rm s}^{-1},\ \ \ {\rm tr}\,\dot{\boldsymbol{\varepsilon}}=0,
\end{equation}
was determined randomly. Instead of system (\ref{eq:1_column}) and (\ref{eq:23_column}), we solve Eq.~(\ref{eq:A_noncons}):
\[\dfrac{{\rm d}\AAA}{{\rm d} t}+\AAA\dot{\boldsymbol{\varepsilon}}=-\dfrac{\PSI}{\tau},\]
where ${\rm d}/{\rm d}t$ is the material time derivative. The numerical solution is depicted in Fig.~\ref{fig:fig2}, where the typical time evolution profiles of $A_{ij}$, $A_{ii}$, $G_{ij}$ and $T^v_{ij}$ for air  with $\tau=1.45\cdot10^{-9}$ s are given.

We note that, in general case, the off-diagonal entries $A_{ij}$, $i\neq j$ of the elastic distortion $\AAA$ do not exhibit  steady-state behavior as shown in Fig.~\ref{fig:fig_air}(a) and (b) for singular values $a_i$ and the shear stress $T^v_{21}$. This is because distortion $\AAA$ (non-symmetric matrix) contains not only the information about deformation of the fluid particle, but $\AAA$ also comprises the information about rotations of the particle after it rearranges with its neighbors (see also (\ref{eq:shear_0}) and a comment immediately after that). In contrast, the deformation tensor $\GG=[G_{ij}]=\AAA^\mathsf{T}\AAA$ does not contain such rotations; it describes solely the deformation of the particle, and hence, all entries $G_{ij}$ of the matrix $\GG$ exhibit a steady-state behavior\footnote{This is true only for the incompressible flows. See the example for a compressible flow below.} similarly to the singular values $a_i$ and the stress $T^v_{12}$.   

Fig.~\ref{fig:fig2} shows that the deformation of fluid particles is preserved, see Figs.~\ref{fig:fig2}(a) and (d), and the shear stresses remain constant, Fig.~\ref{fig:fig2}(c), after time $t=\tau$, but particles continue to rotate at a constant angular velocity; slopes of $A_{ij}$, $i\neq j$ are constant, Fig.~\ref{fig:fig2}(b).

\paragraph{\textit{Remark \arabic{mypar}.}}\stepcounter{mypar} \textit{ As long as a fluid under consideration is simple and a flow is laminar, the information about rotations of fluid particles stored in the distortion $\AAA$ (see Fig.~\ref{fig:fig2}, (b)) can be ignored and, for example, the deformation tensor $\GG=\AAA^\mathsf{T}\AAA$ can be used as the state variable instead of $\AAA$. The situation becomes quite different if we consider complex fluids, \textit{e.g.}, liquid crystals~\cite{Sadovskii2013}, where the orientation of particles plays an important role, and therefore particle rotations cannot be ignored. One may expect that the same is true for turbulent flows of simple Newtonian fluids~\cite{Shliomis1967,Berezin1997}.
} 

Fig.~\ref{fig:fig2}(c) demonstrates that the viscous stresses (black dashed lines)
\begin{equation*}\label{eq:visc_classic}
\TT^{NS}=\eta((\dot{\boldsymbol{\varepsilon}}+\dot{\boldsymbol{\varepsilon}}^\mathsf{T})-\frac{2}{3}{\rm tr}\,\dot{\boldsymbol{\varepsilon}}\II)
\end{equation*}
obtained in the framework of the classical theory of viscous Newtonian fluids are the steady state values for the stresses $\TT^v=-\rho \AAA^\mathsf{T} E^v_{\AAA}$ computed within the framework of the hyperbolic theory. It should be stressed that we do not explicitly prescribe to $\TT^v$ to relax towards $\TT^{NS}$ in a manner of (\ref{eq:Maxwell_idea}), but instead the steady state values of $\TT^v$ give automatically the correct answer if the viscous internal energy $E^v$, the strain dissipation function $\PSI$ and time $\tau$ are properly defined.

\begin{figure}[t]
\includegraphics[trim = -10mm 140mm 0mm 10mm, clip, scale=0.7]{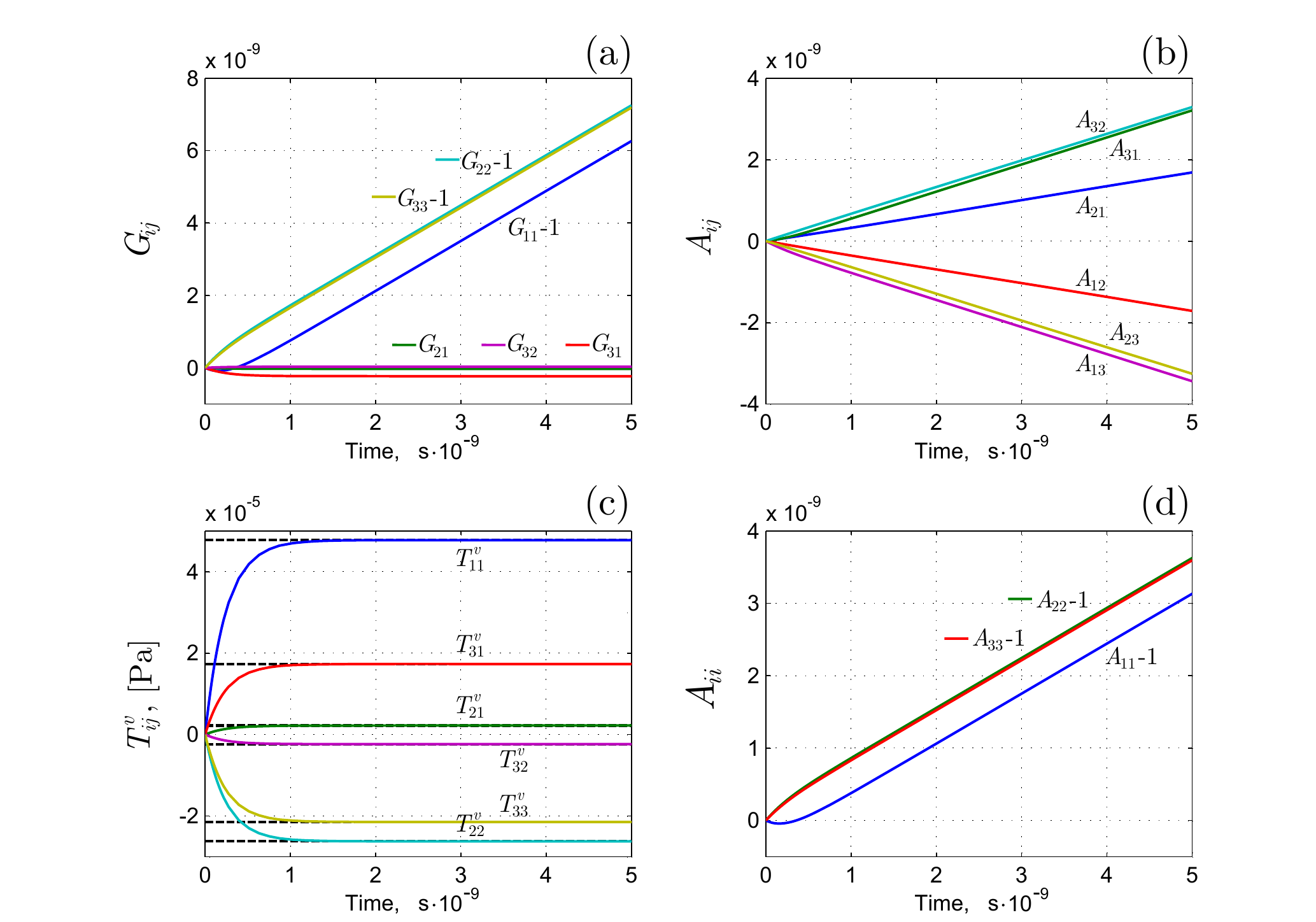}
\caption{\it \footnotesize Results of numerical solution of Eq.~(\ref{eq:A_noncons}) for a compressible flow of air given by arbitrary chosen rate of strain tensor (\ref{eq:example2}). The black dashed lines in (c) are the components of the classical viscous stress tensor $\TT^{NS}$.}\label{fig:fig3}
\end{figure}

Finally, we examine the model in the case of compressible flows. As previously, the rate of strain tensor
\begin{equation}\label{eq:example2}
\dot{\boldsymbol{\varepsilon}}\equiv\left[\dfrac{\partial v_i}{\partial x_j}\right]=\left[
\begin{array}{rrr}
 0.62  &   0.40  &   1.14\\
-0.28  &  -1.41  &   0.59\\
-0.19  &  -0.72  &  -1.28
\end{array}
\right]\ {\rm s}^{-1},\ \ \ {\rm tr}\,\dot{\boldsymbol{\varepsilon}}=-2.07,
\end{equation}
was determined randomly. Results are given in Fig.~\ref{fig:fig3}. In particular, Fig.~\ref{fig:fig3}(c) shows that, for an arbitrary given compressible flow, the classical viscous stresses $\TT^{NS}$ (black dashed lines) are also obtained as the steady state solution to the hyperbolic model.

\begin{table}
\begin{center}
\begin{tabular}[c]{| c | c | c | c | c |}
\hline
      		&						&						&				&						\\
      		&	$\eta$, [Pa$\cdot$s]&	$\rho_0$, [kg/m$^3$]&	$c_s$, [m/s]&	$\tau$, [s]			\\[4mm] \hline
      		&						&						&				&						\\
air   		&	$18.21\cdot10^{-6}$	&	1.2					&	250			& 	$1.457\cdot10^{-9}$	\\[4mm] \hline
      		&						&						&				&						\\
water 		&	$1.002\cdot10^{-3}$	&	1000				&	150			& 	$2.672\cdot10^{-10}$\\[4mm] \hline
      		&						&						&				&						\\
``honey'' 	&	5					&	1420				&	150			& 	$9.390\cdot10^{-7}$	\\[4mm] \hline
\end{tabular}
\end{center}
\caption{\it \footnotesize Values of the settled time $\tau$ for a given shear sound velocity (at rest) $c_s$ for three typical Newtonian fluids.}\label{tab:tab1}
\end{table}

\section{\label{sec:Discuss}Discussion}

\paragraph{Relation to the NSEs.} As was shown previously, Newton's viscous law, and subsequently the NSEs \textit{per se}, is obtained in our hyperbolic theory as a steady-state limiting case if the time of observation and a characteristic time of mechanical fluctuations are sufficiently larger than the settled time $\tau$. However, it is also clear that, in general,  solutions to the hyperbolic model and NSEs may not coincide. For example, one may expect that solutions to these models may diverge for strongly sheared and essentially time-dependent flows.

In addition, the hyperbolic model and the NSEs, which are of parabolic type, are two quite different mathematical objects. For example, it is well known that the type of PDEs influences strongly the choice of a numerical method,  the way of solution of the initial and boundary value problems,~\textit{etc}.

\paragraph{Elastic response of Newtonian fluids. } According to the definition of the PSL time $\tau$, Newtonian fluids can exhibit a pure elastic response if a characteristic time $t\,'$ of shear fluctuations is less than $\tau$. In fact, if the loading reverses its direction of action for time $t\,'<\tau$, then no bond destructions and subsequent particle rearrangements occur,  thus, the fluid cannot flow but exhibits a pure elastic response. This time is, however, extremely small for such fluids as water and air; of order $10^{-9}$~s, which corresponds to a perturbation frequency of order $1$~GHz, and this is far below the time scale of the typical day life. 

\paragraph{Numerical implementation. Time step. } It may seem that such a small time scale for the strain dissipation characteristic time ($\tau \approx 10^{-9}$ s) prevents practical use of the model for such fluids as air or water because the integration time step $\Delta t$ for models of relaxation type should be of the order of $\tau$ or less. However, the situation is not so unfavorable. Because the solutions to (\ref{eq:A_noncons}) exhibit a steady-state behavior in the time interval $\tau<t<t'$, where $t'$ is the moment of time before which the assumption $\boldsymbol{\dot{\varepsilon}}=const$ gives a satisfactory approximation,  $\Delta t$ can be taken of the order of $t'$. Therefore, the task is to invent a procedure that determines the values of $\AAA$ corresponding to this steady-state regime. After that, it remains to compute the viscous stress tensor using (\ref{eq:stress_total2}), (\ref{eq:Tvisc}). Such a strategy is used, for example, in multiphase flow problems~\cite{SaurelStiff2007}.

\paragraph{Lack of divergence form.} One of the drawbacks of the model is that Eq.~(\ref{eq:model_eul_b}) for  elastic distortion is not in a divergence form. A conservative extension of the model based on a continuum notion of flow defects, \textit{i.e.}, microscopic slips of one cluster of particles relative to another, is discussed in~\cite{PeshGrmRom2014}.

\section{Summary}

The conventional continuum interpretation of fluids and solids as a system of material particles has enlarged with the assumption of deformability of the particles. In this settings, we have discussed an alternative to the phenomenological concept of the viscosity coefficient: the particle settled life (PSL) time $\tau$~\cite{Frenkel1955}, also called here the strain dissipation characteristic time. Unlike the viscosity coefficient, the PSL time represents an observable in principle material characteristic with a clear microscopic physical meaning, see Sect.~\ref{sec:solidsfluids}.


The proposed characteristic allows us to develop an intrinsically hyperbolic mathematical model for flows of viscous fluids. The base of our theory was the non-stationary equations of non-linear elastoplasticity derived in~\cite{GodRom1972,God1978,GodRom1998,GodRom2003}. The internal resistance to flow was interpreted as elastic stretching of particle bonds; a flow is a result of particle rearrangements, or strain dissipation, after bonds are destroyed.

The important properties of the discussed model can be summarized as follows. Most of these properties distinguish our approach from the classical theory of viscous Newtonian fluids and its hyperbolic extensions as, for example, Extended Irreversible Thermodynamics~\cite{Jou2010}.
\begin{itemize}
\setlength{\itemsep}{5pt}
\item
 the model is {\it causal}. The shear sound velocities are always finite and less then the longitudinal sound velocity. Moreover, the model is {\it hyperbolic} if the total energy density $\rho\scE$ is a convex function with respect to the state variables, see Sect.~\ref{sec:Hyperb} for a more precise formulation. In other words, it has been demonstrated that there are no physical reasons saying that the viscous Newtonian flows should be exclusively modeled with the parabolic type PDEs, and that it is now possible to consider the whole range of viscous flows (equilibrium or non-equilibrium) from the standpoint of wave propagation physics.
\item
 the model is {\it non-equilibrium} and potentially can be applied to far-from-equilibrium strongly sheared flows, see Sect.~\hyperref[sec:nonequilibr]{2.7}.
\item
 the model is {\it fully thermodynamically consistent}. The fulfilment of the first and second law of thermodynamics is guaranteed by the structure of the governing equations, see Sects.~\ref{sec:time_evol} and \ref{sec:strain_diss}.
\item
the model is suitable for the use of advanced {\it high-accuracy Godunov-type numerical methods}, \textit{e.g.}, see~\cite{BartonRom2010,KulikPogorSemen}. 
\item
the model is applicable to modeling of {\it non-Newtonian flows} provided the strain dissipation time $\tau$ and viscous internal energy $E^v$ are properly defined functions of the state variables $\AAA$, $\rho$, $s$. 
\item
the developed theory provides a unified framework for simulation of flows of viscous fluids as well as of irreversible deformation in solids. The ideal cases as inviscid fluids, elasticity, or ideal plasticity are obtained in this theory as the limiting cases, see Sect.~\ref{sec:solidsfluids}.
\end{itemize}

Finally, we examined the theory for the simple shear flows and arbitrary incompressible as well as compressible flows of Newtonian fluids, such as air, water, and honey, and demonstrated that Newton's viscous law can be recovered in the framework of the hyperbolic theory as the steady-state limit. Basic relations~(\ref{eq:tau_cs}) and (\ref{eq:eta_tau}) for the viscosity coefficient, strain dissipation time $\tau$, and shear sound velocity were obtained.

\section*{Acknowledgement}
This work would be impossible without thoughtful and fruitful discussions with Professor Miroslav Grmela during the post-doctoral studies of I.P. at the Ecole Polytechnique de Montreal. I.P. acknowledges financial support from the Labex MEC (ANR-10-LABX-0092) and A*MIDEX project (ANR-11-IDEX-0001-02), funded by the ``Investissements d'Ave\- nir'' French Government program managed by the French National Research Agency (ANR). 
E.R.~acknowledges financial support from the Russian Foundation for Basic Research (grants 13-05-00076, 13-05-12051) and the Siberian Branch of Russian Academy of Sciences (Integration Project No~127).

\bibliographystyle{vancouver}
\bibliography{Bibliography}

\end{document}